\documentclass[
twocolumn,%
nofootinbib,%
superscriptaddress,%
amsfonts,amsmath]{revtex4}

\usepackage{amsmath}
\newcommand{\be}{\begin{equation}}
\newcommand{\ee}{\end{equation}}
\newcommand{\bea}{\begin{eqnarray}}
\newcommand{\eea}{\end{eqnarray}}
\newcommand{\non}{\nonumber}
\usepackage{graphicx}	\begin{document}

\title{Gray-body factor and infrared divergences in 1D BEC acoustic black holes}

\bigskip

\author{Paul~R.~Anderson}
\email{anderson@wfu.edu}
\affiliation{Department of Physics, Wake Forest University, Winston-Salem, North Carolina 27109, USA}
\author{Roberto~Balbinot}
\email{Roberto.Balbinot@bo.infn.it}
\affiliation{Dipartimento di Fisica dell'Universit\`a di Bologna, Via Irnerio 46, 40126 Bologna, Italy}
\affiliation{INFN sezione di Bologna, Via Irnerio 46, 40126 Bologna, Italy}
\affiliation{Centro Studi e Ricerche E. Fermi, Piazza del Viminale 1, 00184 Roma, Italy}
\author{Alessandro~Fabbri}
\email{afabbri@ific.uv.es}
\affiliation{Centro Studi e Ricerche E. Fermi, Piazza del Viminale 1, 00184 Roma, Italy}
\affiliation{Dipartimento di Fisica dell'Universit\`a di Bologna, Via Irnerio 46, 40126 Bologna, Italy}
\affiliation{Departamento de F\'isica Te\'orica and IFIC, Universidad de Valencia-CSIC, C. Dr. Moliner 50, 46100 Burjassot, Spain}
\affiliation{Laboratoire de Physique Th\'eorique, CNRS UMR 8627, B\^at. 210, Universit\'e Paris-Sud 11, 91405 Orsay Cedex, France}
\author{Renaud~Parentani}
\email{renaud.parentani@th.u-psud.fr}
\affiliation{Laboratoire de Physique Th\'eorique, CNRS UMR 8627, B\^at. 210, Universit\'e Paris-Sud 11, 91405 Orsay Cedex, France}

\begin{abstract}
It is shown that the gray-body factor for a one-dimensional elongated Bose-Einstein condensate (BEC) acoustic black hole with one horizon
does not vanish in the low-frequency ($\omega\to 0$) limit. This implies that the analog Hawking radiation is dominated by the emission
of an infinite number ($\frac{1}{\omega }$) of soft phonons
in contrast with the case of a Schwarzschild black hole where the gray-body factor vanishes as $\omega\to 0$ and the spectrum is not
dominated by low-energy particles.  The infrared behaviors of certain correlation functions are also discussed.
\end{abstract}


\maketitle

One of the most exciting results of modern theoretical physics is the prediction made by  Hawking in 1974 \cite{Hawking:1974sw} that black holes are not `black', but should emit particles with a thermal spectrum at a temperature
\be\label{teha}
T_H=\frac{\hbar}{8\pi GMk_B}  \;, \ee
where $M$ is the mass of the black hole.
Unfortunately, at present, an experimental verification of this emission seems out of reach: the emission temperature for a solar mass black hole (BH) is of the order of $10^{-7}\ K$. For this reason growing interest has been manifested in recent years on analog black holes consisting of condensed matter systems that are expected to show phenomena analogous to Hawking  radiation \cite{unruh}.
Of these, Bose-Einstein condensates (BECs) provide one of the most promising settings for the experimental detection of   these effects~\cite{garayetal, fnum}.

The Hawking effect for BHs in asymptotically flat spacetimes engenders an interesting interplay between thermal effects, infrared divergences, and gray-body factors.
First one should note that a Planckian distribution for the number of created particles has an infrared divergence with the result that the spectrum is dominated by low energy
particles.  However, the Planckian distribution is filtered by a ``gray-body'' factor $\Gamma^{(j)}(\omega)$ due to an effective potential which takes into account the scattering of the particles by the spacetime geometry.  The potential has the shape of a barrier whose height increases with the angular quantum number $l$, so that the emission is dominated by (massless) particles in the $l=0$ mode \cite{page}.

The number of particles emitted at frequency $\omega $ and quantum number $j$ is
\be \label{gbp}
N_\omega^{(j)}=\frac{\Gamma^{(j)}(\omega)}{e^{\frac{\omega}{k_BT_H}}-1}\ .\ee
For BHs in asymptotically flat spaces at low $\omega$ the characteristic leading order behaviour is
\be\label{gblf}
\Gamma_\omega^{(j)}\sim A_H \omega^2\ , \ee
where $A_H$ is the area of the BH horizon. Because of this, low energy modes are suppressed    and the gray-body factor regularizes the infrared
divergence ($1/\omega$) of the Planckian distribution.

   In this paper we calculate the low frequency behaviour of the gray-body factors for  BEC acoustic BHs   and show that they do not remove the infrared divergences of the Planckian distribution.  We also investigate the question of whether and under what circumstances infrared divergences occur in certain correlation functions for these models.

Following a by now standard procedure, we begin by splitting the fundamental bosonic operator for the atoms, $\hat \Psi$, into a $c$-field part, $\Psi_0$, which describes the condensate in the mean field
approximation, and an operator part, $\hat\phi$, which describes the quantum fluctuations about the mean. $\Psi_0$ satisfies the Gross-Pitaevski equation, while
$\hat\phi$ satisfies the Bogoliubov-de Gennes equation. Using a density-phase representation for $\hat\Psi$ \cite{blv}
\be\label{dp}
\hat\Psi=\sqrt{n+\hat n_1}e^{i(\theta+\hat\theta_1)}
\ee
the fluctuations equation can be written as\footnote{
This approximation is valid in a regime, 
denoted as `1D mean field' in \cite{mest}, where the
system is accurately described by a single order parameter obeying an effective 1D Gross-Pitaevskii equation.}
\bea
\hbar \partial_t\hat\theta_1 &=& -\hbar\vec v_0\vec\nabla\hat\theta_1-\frac{mc^2}{n}\hat n_1+\frac{mc^2}{4n}\xi^2\vec\nabla [n\vec\nabla (\frac{\hat n_1}{n})],\label{dpeqs1} \\
\partial_t\hat n_1 &=& -\vec\nabla (\vec v_0 \hat n_1 + \frac{\hbar n}{m}\vec\nabla \hat \theta_1 ) \label{dpeqs2}
\eea
where $\vec v_0=\frac{\hbar \vec\nabla \theta}{m}$ is the condensate velocity, $n=|\Psi_0|^2$ the condensate density, $c\equiv \sqrt{\frac{ng}{m}}$ the speed of sound, $g$ the atomic interaction coupling, and $\xi=\frac{\hbar}{mc}$ the healing length.

On scales much larger than $\xi$ one can neglect the last term in (\ref{dpeqs1}) which then becomes
\be \label{dphy}
\hat n_1=-\frac{\hbar n}{mc^2}[\vec v_0 \vec \nabla \hat \theta_1+\partial_t\theta_1]\ .
\ee
This is the so called hydrodynamical approximation. Inserting Eq. (\ref{dphy}) into Eq. (\ref{dpeqs2}) one gets a decoupled equation for the phase fluctuations
\be\label{phhy}
-(\partial_t+\vec\nabla\vec v_0)\frac{n}{mc^2}(\partial_t+\vec v_0\vec\nabla)\hat\theta_1 + \nabla (\frac{n}{m}\vec\nabla\hat\theta_1)=0\ .\ee
This equation can be rewritten as a covariant
equation
\be \label{kg}
\Box\hat\theta_1=0\ee
in a fictitious curved four-dimensional space-time with the following metric
\begin{equation}\label{acm}
g_{\mu\nu}=\frac{n}{mc}\left(
  \begin{array}{cccc}
   -(c^2-\vec v_0^2) & -v_0^i \\
     -v_0^j & \delta_{ij} \\
  \end{array}\right).
\end{equation}
The covariant d'Alembertian operator is
\be \label{do}
\Box\equiv \frac{1}{\sqrt{-g}}\partial_{\mu}(\sqrt{-g}g^{\mu\nu}\partial_{\nu}) \;,
\ee
where $g\equiv det g_{\mu\nu}$.

For the sake of simplicity we shall make a set of assumptions (see \cite{fnum}). First we assume that the condensate is infinite and elongated along the $x$ axis with transverse
size $l_{\perp}$ constant and much smaller than $\xi$. So the dynamics is frozen along the transverse direction and the system becomes effectively one dimensional. We further assume that the flow is stationary and directed along $x$ from right to left with a constant velocity, i.e. $\vec v_0=-v\hat x$, with $v$ a positive constant.
This implies that the density $n$ of the atoms is also constant.
Nontrivial configurations are obtained by allowing $g$ and hence the sound speed $c$ to vary with $x$. The profile $c(x)$ is chosen so that $c>v$ for $x>0$ and $c<v$ for $x<0$. We have therefore a supersonic region ($x<0$) separated at $x=0$ from a subsonic one ($x>0$). This configuration describes a so called ``acoustic BH'' and $x=0$ (where $c=v$) is the sonic horizon. The profile
$c(x)$ is assumed to vary smoothly (i.e. on scales $\gg \xi$) from an asymptotic value $c_L$ ($<v$) for $x\to -\infty$ to $c_R$ ($>v$) for $x\to +\infty$).

To proceed with quantization, we neglect the transverse modes and expand $\hat \theta_1$ using a basis constructed from mode solutions to
\be \label{kgmode}
\Box \psi(t,x)=0\ . \ee
It is useful to rescale the modes so that
\be \label{dr}
\psi=\sqrt{\frac{mc}{n\hbar l_{\perp}^2}}\ \chi  \;, \ee
and then to rewrite (\ref{kgmode}) as
\be \label{fe2d}
(\Box^{(2)}-V)\chi(t,x)=0 \;,
\ee
where $\Box^{(2)}$ is the covariant d'Alembertian associated with the two-dimensional ($t,x$) section of the acoustic metric (\ref{acm}) and
\be\label{pot}
V\equiv- \frac{1}{2}\frac{d^2c}{dx^2}(1-\frac{v^2}{c^2})+\frac{1}{4c}[1-\frac{5v^2}{c^2}](\frac{dc}{dx})^2 \ .\ee
In order to find the solutions to this equation, we apply two cordinate transformations.
First we introduce a ``Schwarzschild'' time $t_s$ as
\be \label{sct}
t_s=t-\int^{x}dy\frac{v}{c^2(y)-v^2}\ee
and then a ``tortoise'' spatial coordinate $x^*$ as
\be\label{toco}
x^*=\int^x dy \frac{c(y)}{c^2(y)-v^2}\ .\ee
The second one maps the subsonic region $0<x<\infty$ to $-\infty<x^*<+\infty$, i.e. the horizon corresponds to the asymptote $x^*\to -\infty$.\footnote{In this paper we concentrate only on the region exterior to the horizon, the subsonic ($x>0$) one. A similar analysis can be performed in the interior, supersonic, $x<0$ region.}
The utility of these transformations is to bring the mode equation into the simple form
\be\label{mesf}
\left(-\frac{\partial^2}{\partial t_s^2}+\frac{\partial^2}{\partial x^{*2}}-V_{eff}\right)\chi =0 \ee
where $V_{eff}=\frac{c^2-v^2}{c}V$.
We look for stationary solutions
\be \label{stso}
\chi=e^{-i\omega t_s}\chi_\omega(x^*)=e^{-i\omega t}\varphi_\omega (x)\ee
where $\varphi_\omega$ is the spatial part of the mode function in the original coordinates (\ref{acm}) and  $\chi_\omega$ satisfies
\be\label{weq}
\left(\omega^2+\frac{\partial^2}{\partial x^{*2}}-V_{eff}\right)\chi_\omega=0.\ee
In the asymptotic regions, $x^*\to \pm\infty$, $V_{eff}$ vanishes and the solutions of (\ref{mesf}) are simply plane waves
\be\label{assol}
\chi \sim e^{-i\omega(t_s\pm x^*)}=e^{-i\omega[t\mp\int^x \frac{dy}{c\mp v}]}.\ee

A complete basis for the solutions of (\ref{mesf}) is formed by two sets of modes, $\chi_I$ and $\chi_H$. These are easily pictured in the diagrams of Figs. 1 and 2 representing the
causal structure (Penrose diagram) of the exterior (subsonic) region. Details on how to construct such diagrams can be found in \cite{Barcelo:2004wz}.
The modes $\chi_I$ originate at past null infinity ($I^-$) and because of the potential term in Eq.\ (\ref{mesf}) are partially transmitted towards the future horizon ($H^+$) and partially reflected to future null infinity ($I^+$), see Fig. 1.
\begin{figure}[b]
\includegraphics[scale=0.6]{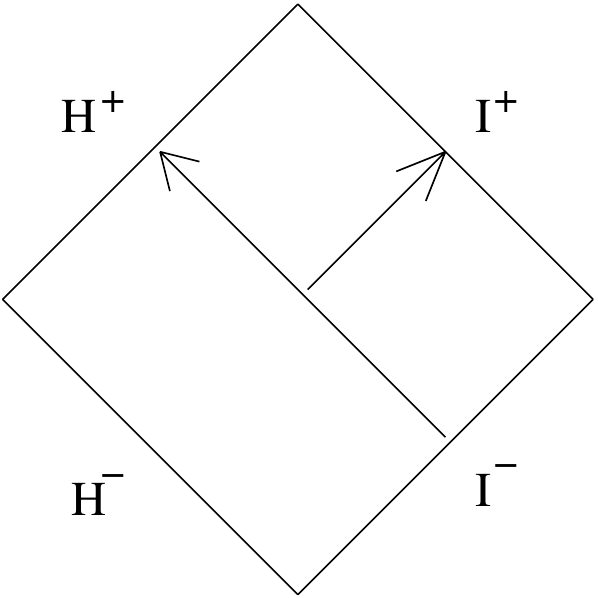}
\caption{Modes $\chi_I$ originating from $I^-$ and transmitted (reflected) to $H^+$ ($I^+$).}
\label{fig:1}
\end{figure}
The modes $\chi_H$ can be thought of as originating on the past horizon ($H^-$) of the analytically extended manifold of the effective space-time. They are partially transmitted to future  null infinity ($I^+$) and partially reflected to the future horizon ($H^+$), see Fig. 2.
\begin{figure}[b]
\includegraphics[scale=0.6]{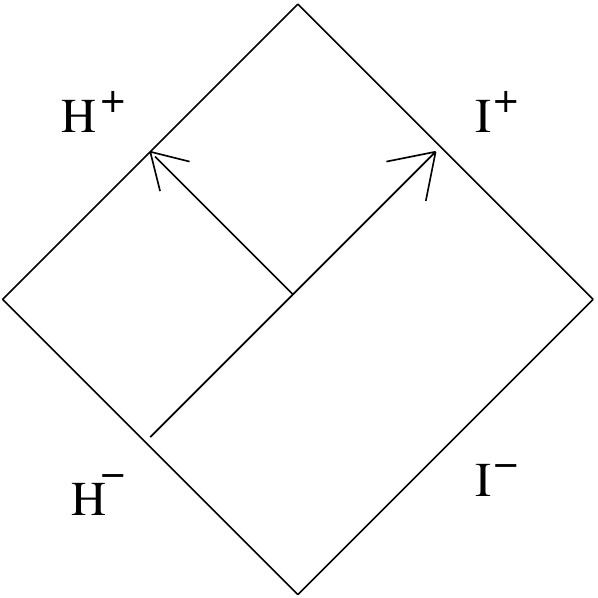}
\caption{Modes $\chi_H$ originating from $H^-$ and transmitted (reflected) to $I^+$ ($H^+$).}
\label{fig:2}
\end{figure}
More specifically
\bea \label{chiI}
\chi_I &=& \frac{1}{\sqrt{4\pi \omega}}e^{-i\omega t_s}\chi_\omega^I\ ,\\
\label{chiH}
\chi_H &=& \frac{1}{\sqrt{4\pi \omega}}e^{-i\omega t_s}\chi_\omega^H\ ,\eea
where $\chi_\omega^I$ and $\chi_\omega^H$ are solutions of Eq. (\ref{weq}) with the following asymptotic behaviours
\begin{equation}\label{aschiI}
\chi_I=\Big\{
  \begin{array}{c}
   e^{-i\omega(t_s+x^*)}+R_I(\omega) e^{-i\omega(t_s-x^*)}, \ \ x^*\to +\infty \\
   T(\omega) e^{-i\omega(t_s+x^*)} , \ \  x^* \to -\infty \\
   \end{array}
   \end{equation}
and
\begin{equation}\label{aschiH}
\chi_H=\Big\{
  \begin{array}{c}
   T(\omega) e^{-i\omega(t_s-x^*)}, \ \ x^*\to +\infty \\
   e^{-i\omega(t_s-x^*)}+R_H(\omega) e^{-i\omega(t_s+x^*)} , \ \  x^* \to -\infty  \\
   \end{array} \ .
   \end{equation}
    The two reflection coefficients satisfy $|R_I(\omega)|^2=|R_H(\omega)|^2$ and also the unitary relation $|R_H(\omega)|^2+|T(\omega)|^2=1$.
  The gray-body factor we are looking for is $\Gamma=|T(\omega)|^2$.
   It represents the probability that a phonon
    originating from the past horizon reaches future null infinity.
   This is also equal to the absorption probability
  of an ingoing phonon from past null infinity \cite{dewitt}.

  To compute $T(\omega )$ we shall employ a very simple, although not general, method (see for example \cite{kantietal}) which consists in solving Eq. (\ref{weq}) for $\chi_\omega$ in the infrared limit ($\omega \to 0$) for finite $x^*$, taking the limit $x^*\to\pm\infty$ and matching the solution there with the asymptotic forms (\ref{aschiI}) and (\ref{aschiH}) developed for small $\omega $.
  For fixed $x^*$ and in the limit $\omega \to 0$, we can neglect the first term in Eq. (\ref{weq}) which can then be rewritten, in terms of the original variable $x$, as
  \be\label{wez}
  \partial_x[\frac{(c^2-v^2)}{c^2}\partial_x (\sqrt{c}\chi_0)]=0\ .\ee
  This can be immediately integrated, giving
  \bea \label{chi0}  \chi_0&=& \frac{c_2}{\sqrt{c(x)}}+\frac{c_1}{\sqrt{c(x)}}\int^x dy\frac{c^2(y)}{c^2(y)-v^2}\non \\
  &=& \frac{c_2}{\sqrt{c(x)}}+\frac{c_1}{\sqrt{c(x)}}\int^{x^*}c(y^*)dy^* \ ,\eea
  where $c_{1,2}$ are integration constants.
   From this we can extract the two asymptotic limits
  \bea  \label{chi0hor} && \chi_0 \to \frac{c_2}{\sqrt{v}}+c_1\sqrt{v} \ x^*\ ,\ x^*\to -\infty\ , \\
  \label{chi0inf} && \chi_0 \to \frac{c_2}{\sqrt{c_R}} + c_1 \sqrt{c_R} \ x^* \ , \ x^*\to +\infty \ .  \eea
  These behaviours should then be compared with the small $\omega $ expansion of the spatial part of (\ref{aschiH})
  \bea
  \label{horH}
 &&  \chi^{H }_{\omega\to 0} \to 1+R_H +i\omega (1-R_H)x^*\ , x^*\to -\infty, \\ \label{infH}
 &&  \chi^{H }_{\omega \to 0} \to  T + i\omega Tx^*\ , x^*\to +\infty \ .\eea
 Equating eq. (\ref{chi0hor}) with (\ref{horH}) and (\ref{chi0inf}) with (\ref{infH}) we get
 \bea \label{maeq}
 \label{a}\frac{c_2}{\sqrt{v}} &=&  1+R_H\ , \\
 \label{b}c_1\sqrt{v} &=& i\omega (1-R_H)\ , \\
 \label{c}\frac{c_2}{\sqrt{c_R}} &=& T\ , \\
 \label{d}c_1\sqrt{c_R} &=& i\omega T\ .\eea
 Dividing (\ref{a}) by (\ref{b}) and  (\ref{c}) by (\ref{d})  one finds $R_H=\frac{c_R-v}{c_R+v}$ from which
 \be \label{res}
 |T|^2 = 1-|R_H|^2=\frac{4c_Rv}{(c_R+v)^2}\ .\ee
 This shows that the gray-body factor for a 1D acoustic BH for the realistic profile $c(x)$ does not vanish in the $\omega \to 0$ limit.\footnote{In the limit $\omega \to 0$, the reflection and transmission coefficients
only depend on the asymptotic values of $v$ and $c$, and,
  in this limit, the S-matrix possesses the same form as that found in
Section IV.A of ~\cite{Fabbri:2010tj} for a step-like discontinuity in the sound velocity profile,  with the replacement $c_R\to c_r$ and $v\to c_l$.}
 This conclusion explains the results of the numerical analysis of \cite{Macher:2009nz}, see Figures 11 and 13.
 Interestingly, (\ref{res}) radically differs from the standard result found for asymptotically flat 4D BHs,
 for which $|T|^2\propto \omega^2$~\cite{page, uno, due}.

 A non vanishing gray-body factor in the infrared limit is however not peculiar to acoustic BHs.  One has been found~\cite{kantietal} for the $l=0$ mode of a massless minimally coupled scalar field in  Schwarzschild-de Sitter spacetime (SdS) which is a solution to Einstein's equations for a BH immersed in de Sitter space.  The gray-body factor is
 \be \label{ressds}
 |T|^2_{SdS}=\frac{4r_C^2r_H^2}{(r_C^2+r_H^2)^2} \;,
 \ee
  which is quite similar to the result~\eqref{res}.
 Here $r_H$ is the radius of the BH horizon and $r_C$ the radius of the cosmological horizon. The finite region between the two horizons $[r_H,r_C]$ is mapped, by a tortoise-like coordinate $x^*$, to $-\infty<x^*<+\infty$ as in BECs.\footnote{Even though the size is finite in the $r$ coordinate, the modes oscillate an infinite number of times before reaching the horizons, so it is as if the length of the box is infinite on both ends.}
 The two expressions (\ref{res}) and (\ref{ressds}) are mapped into each other by the substitution $r_H\leftrightarrow \frac{1}{\sqrt{v}},\ r_C\leftrightarrow \frac{1}{\sqrt{c_R}}$.
 This is not surprising since when performing a dimensional reduction along the transverse angular variables ($\theta,\ \phi$) for the $l=0$ spherically symmetric component one gets
 $4\pi r^2$ as the area of the transverse space, whereas in the acoustic BH, due to the conformal factor present in the acoustic metric, the transverse area is $\frac{n}{mc}l_{\perp}^2$.
 This explains the correspondence $r^2\leftrightarrow \frac{1}{c}$ in the term $\sqrt{-g}$ entering the d'Alembertian operator.

\noindent  The existence of a nonvanishing infrared limit for the gray-body factor in the Schwarzschild-de Sitter case was attributed in~\cite{kantietal} to the finite size of the $[r_H,r_C]$ region in which the propagation of the modes was considered. As we have seen the same result is obtained in the infinite space of our 1D acoustic BHs  with just one horizon. The feature that these acoustic BHs and SdS spacetimes share is the existence of an everywhere bounded (not diverging) solution of the $\omega\to 0$ equation (\ref{wez}), namely the first term in Eq. (\ref{chi0}). 
For acoustic BHs, this solution corresponds, in terms of the original field [see Eq. (\ref{dr})], to a classical constant field solution (for the SdS case see \cite{duerefs}). 
For both SdS and Schwarzschild BHs the corresponding term in 
Eq. (\ref{chi0}) is proportional to $r$. Thus it is bounded in the SdS case where  $r_H<r<r_C$, but it is unbounded for the Schwarzschild case where $r_H<r<\infty$.

 In view of our result, we can conclude that, unlike the standard Schwarzschild BH, the Hawking-like emission in a 1D acoustic BH is dominated by soft phonons, since the number of such particles (see Eq. (\ref{gbp})) diverges in the infrared limit.  Thus the gray-body factor no longer cancels the $\frac{1}{\omega}$ divergence of the Planckian distribution factor. However, from an experimental point of view these emitted phonons may be difficult to detect.

The gray-body factor also affects the IR behaviour of correlation functions.
As shown in~\cite{Balbinot:2007de}, a more promising way of observing  the signal of Hawking radiation in a BEC is through the density density correlation function
\be\label{dde}
G_2(t,x;t,x')=\lim_{t\to t'} \langle \hat n_1(t,x)\hat n_1(t',x')\rangle \;, \ee
which in the hydrodynamical approximation can be written, using Eq. (\ref{dphy}), as
\be \label{ddehy}
G_2(t,x;t,x')=A\lim_{t\to t'}D[\langle \{\hat \theta_1(t,x)\hat \theta_1(t',x')\}\rangle]  \;.
\ee
where $\{ , \}$ stands for the anticommutator, 
$A=\frac{\hbar^2 n^2}{2m^2c^2(x)c^2(x')}$ , and the differential operator $D$ is
\be\label{dop}
D=\partial_t\partial_{t'}-v\partial_t\partial_{x'}
-v\partial_{t'}\partial_x+v^2\partial_x\partial_{x'} \ \;. \ee
 The expectation value is taken in the Unruh state,
  which is
  the quantum state that describes Hawking emission of thermal phonons.\footnote{See \cite{abfp} for more details on the correct choice of quantum state after the formation of a sonic BH.
 Note also that the notation in this paper differs from that in~\cite{abfp}.  One can change to that notation by letting $\chi_H \rightarrow u^R_H$, $\chi_I \rightarrow u^R_I$,
   $\chi^H_\omega \rightarrow \chi^R_H$ and $\chi^I_\omega \rightarrow \chi^R_I$. }
 By restricting to points outside the horizon it takes the form (see (\ref{dr}))
\be \langle \{ \hat{\theta}_1(t,x), \hat{\theta}_1(t',x') \} \rangle = \frac{m}{n \hbar \ell_\perp^2} \sqrt{c(x) c(x')} ( I  + J)  \;, \label{IplusJ} \ee
where
\bea
 & &  I =  \int_0^\infty d \omega \frac{\left[\chi_H(\omega,t,x) \, \chi_H^*(\omega,t',x') +c.c.\right]}{\sinh\left(\frac{\pi \omega}{\kappa}\right)}\ ,
 \label{Idef} \\
 & &  J = \int_0^\infty d \omega \, \left[ \chi_I(\omega,t,x) \, \chi_I^*(\omega,t',x')
   +  c.c. \right] \;. \label{Jdef}
\eea
 Here $\kappa=\frac{1}{2v}\frac{d(c^2-v_0^2)}{dx}|_{hor}=\frac{dc}{dx}|_{hor}$ is the   surface gravity of the horizon for the acoustic metric (\ref{acm}). 
Note that because of the non vanishing of $T$ in the low-frequency limit the infrared behaviour of the expectation value
goes like $\int \frac{d\omega}{\omega^2}$ for large $x$ and $x'$ (coming from (\ref{Idef}));  one factor of $\frac{1}{\omega}$ is due to the usual vacuum term, the additional factor of $\frac{1}{\omega}$ comes from the Planckian distribution factor of the Unruh state. We have numerical evidence \cite{paul} that the same IR divergence persists for any value of $x$ or $x'$.
The same factor of $\frac{1}{\omega^2}$ is present in the two-point function for the $l=0$ mode of a massless minimally coupled scalar field in the Unruh state for both Schwarzschild and SdS black holes. In the Schwarzschild case the infrared divergence is removed, at large distances, by the gray-body factor  (\ref{gblf}) for the modes, and one can show that this happens also at the horizon where the asymptotic behaviours are given in (\ref{aschiI}) and (\ref{aschiH}), and $R\to -1 +O(\omega)$ \cite{longpaper}. For SdS the gray-body factor approaches a constant at low frequency, see (\ref{ressds}), and is similar to the the BEC case.

Despite this fact, a careful analysis of the solutions to the mode equations in the BEC case
shows that when the operator $D$ acts on the expectation value it always brings down two factors of $\omega$ thus removing the infrared divergence and making $G_2$ infrared 
finite.\footnote{ Given the factor $\sqrt{c(x) c(x')}$ in~\eqref{IplusJ} this is a somewhat surprising result which was missed in the analysis of~\cite{abfp} where it was stated
that the infrared divergence in the two-point function results in infrared divergences in the density-density correlation function.}
For finite nonzero values of $x$ this has been seen numerically~\cite{paul}.  It can also be seen analytically by approximating $\chi_\omega$, at low-frequency,  with {\bf $\chi_0$}, where $\chi_0$ is given in (\ref{chi0}) and $c_1$ and $c_2$ are obtained from
(\ref{a})-(\ref{d})
for $\chi^H_\omega$ and by an analogous set of equations for $\chi^I_\omega$. At the horizon, where this approximation is not valid,
we find, for small $\omega$ \cite{longpaper}
 \bea &&\sqrt{c} \chi_H \sim \sqrt{v_0}\left[ e^{-i\omega (t_s-x^*)}+R_H e^{-i\omega (t_s+x^*)}\right] (1+O(\omega x)) \nonumber \ \ \ \ \\
 &&\sqrt{c}\chi_I \sim \sqrt{v_0}Te^{-i\omega (t_s+x^*)}(1+O(\omega x)) \ . \nonumber \eea
A similar analysis shows that the infrared divergence is also removed in the point-split stress-energy tensor for a massless minimally coupled scalar field in the Schwarzschild-de Sitter case (the details for both of these cases will be given in \cite{longpaper}).

Finally, we mention that for profiles for which $V_{eff}=0$ the solutions (\ref{aschiI}) and (\ref{aschiH}), with $T=1$ and $R=0$ are exact and in this case both phase-phase (\ref{IplusJ}) and density-density (\ref{ddehy}) correlation functions are infrared divergent. For these profiles, however, the conformal factor in the metric (\ref{acm}), $\frac{n}{mc}$, goes as $x^{*2}$.  Thus it diverges both at the horizon and at infinity and does not represent physically interesting situations.

{\em Acknowledgments} We thank Iacopo Carusotto, Xavier Busch and Florent Michel for useful discussions. This work was supported in part by the National Science Foundation under Grant Nos. PHY-0856050 and PHY-1308325.

\end{document}